# A comparative, multiscalar, and multidimensional study of residential segregation in seven European capital cities

**Ana Petrović**

Delft University of Technology, Faculty of Architecture, Department of Urbanism

P.O. Box 5043, 2600 GA Delft, The Netherlands

e-mail: a.petrovic@tudelft.nl

**Maarten van Ham**

Delft University of Technology, Faculty of Architecture, Department of Urbanism

P.O. Box 5043, 2600 GA Delft, The Netherlands

e-mail: m.vanham@tudelft.nl

**David Manley**

University of Bristol, School of Geographical Sciences

University Road, Clifton, Bristol, BS8 1SS, UK

Delft University of Technology, Faculty of Architecture, Department of Urbanism

P.O. Box 5043, 2600 GA Delft, The Netherlands

e-mail: d.manley@bristol.ac.uk

**Tiit Tammaru**

University of Tartu, Department of Geography

Vanemuise 46, Tartu 51014, Estonia

e-mail: tiit.tammaru@ut.ee

# A comparative, multiscalar, and multidimensional study of residential segregation in seven European capital cities

**Abstract**

There are relatively few comparative cross-European studies on segregation, and those that do exist often use a single measure of segregation at a single spatial scale. This paper investigates ethnic segregation in seven European capitals – Amsterdam, Berlin, Lisbon, London, Madrid, Paris, and Rome – using the five dimensions of segregation (centralisation, evenness, exposure, clustering, and concentration) at multiple spatial scales. For each dimension, we found very different levels of segregation. Moreover, the impact of scale was different in both between and within cities relative to their cores and hinterlands. Crucially, we found that segregation does not necessarily decrease with spatial scale.

## 1 Introduction

Europe has long been a continent of internal and external migration. In the recent decade, the distribution of migrants from outside Europe, especially asylum seekers, is a major challenge for European unity (see for instance Gessler et al., 2021), and within European countries the distribution of migrants presents many challenges, especially within larger urban areas. Overall levels of ethnic residential segregation in European cities show stark differences between countries and between cities within countries (Arbaci, 2007). A recent comparative study (Tammaru et al., 2016) on socio-economic segregation in European capital cities has reported links between high levels of segregation and increased immigration to Europe. Both ethnic and socio-economic segregation can have important impacts on individuals and on the functioning of cities as integral systems. For example, exposure to other people can affect a wide variety of individual outcomes, such as learning languages, or school or labour market outcomes (Van Ham et al, 2012), which are some of the most crucial factors of the integration of migrants in EU cities.

Despite the importance of ethnic residential segregation and the differences between European countries, there are few comparative, cross-European, studies on the topic (e.g. Andersson et al. 2018; Marcińczak et al. 2021). When segregation has been investigated, it is often done with a focus on a single city or country, overlooking the potential of comparing levels of segregation between different countries. This study investigates ethnic segregation in seven European capitals using an approach which allows direct comparisons of segregation levels between these cities. As international comparisons of segregation are difficult because the size of administrative areas differs, we adopt a multiscale approach, using the same small spatial building blocks. This approach explicitly acknowledges that the degree to which a city appears to be segregated (or not) is, in part, dependent of the scale at which segregation in a city is explored (Sleutjes et al, 2018; Marcińczak et al. 2021). Our bespoke multi-scale measures of population are defined for each residential location and are sensitive to urban form (Petrović et al., 2018), and allow us to take a comparative perspective and investigate at which spatial scales segregation manifests itself in different urban contexts in different countries.

The few existing comparative studies of segregation restrict their measures of segregation to the dissimilarity index. However, segregation can take many forms, and different measures might lead to different outcomes for the same city. Therefore we measure segregation for all five dimensions proposed by Massey and Denton (1989). We contribute to the emerging body of comparative segregation research by using these multiple dimensions of segregation across multiple spatial scales, which allows us to bring different perspectives together to explicitly recognise that the meaning of segregation varies between cities. For instance, a city with small concentrations of immigrants scattered all across the city will have different potential exposures to others compared to a city where immigrant concentrations are highly clustered in specific neighbourhoods. The potential for exposure to 'others' matters, because where exposure is low, the chances for population groups to mix, meet and interact is reduced. Also, when exposure is low, it becomes harder for immigrants to learn the language or to access the local labour market (Danzer et al., 2018). Using multiple dimensions over multiple scales will ensure a better understanding of segregation, and allow a better comparison of segregation levels in different cities.

This paper studies ethnic residential segregation by combining different dimensions of segregation, measured at a range of different spatial scales, for seven European capital cities covering multiple immigration and welfare contexts: Amsterdam, Berlin, Lisbon, London, Madrid, Paris, and Rome. In doing so we address three research questions: First, what are the levels of ethnic segregation in each city and how do these levels vary between the cities? Second, how does segregation manifest itself at different geographical scales, and how does this vary between the cities? Third, how do levels of segregation vary across different dimensions? We use Massey and Denton's (1988) five dimensions of segregation, namely centralisation, evenness, exposure, clustering, and concentration, and to investigate the effects of scale, we use 101 increasingly large bespoke areas (see Petrović et al., 2018). Such bespoke measures more closely represent an individual's residential context than administrative units. While many segregation studies focus on core city areas alone, we consider the entire urban regions of the seven capitals, drawing on the OECD (2012) Functional Urban Areas (FUA).

# 2 Segregation as a multidimensional and multiscale phenomenon

## 2.1 Migration and residential segregation in Europe

Most segregation studies which compare cities were conducted in the US (see, e.g. Krupka, 2007; Rey et al., 2021). Restricting analysis to a single country enables direct comparisons, because ethnic groups are defined in the same way in all cities. Comparative studies in Europe are sparser and include limited numbers of cities, but they demonstrate that ethnic segregation varies across cities and countries (Andersson et al. 2018; Marcińczak et al. 2021). Evidence of high ethnic segregation can be found in studies for some European cities, notably London (Johnston et al., 2016), Amsterdam (Boterman et al., 2021) and Paris (Préteceille, 2011). Southern European cities are less often studied, but have lower levels of ethnic segregation and more complex spatial distributions of ethnic minorities, with greater peripheralisation (or suburbanisation; Malheiros, 2002, Arbaci, 2008). Different levels of segregation are related to the characteristics of the labour and housing markets and welfare systems, as well as with the immigration histories and the overall proportions of ethnic minorities in different countries and cities. Our study presents a mixture of places and histories: each of the countries has a different migration history, which was characterised (in various periods of time) by migration from former colonies, other non-European countries, or within-Europe migration. The reasons for migration also vary, including economic, political, social and cultural reasons.

After the Second World War, north-Western European countries were economically booming, attracting many migrant workers, mostly from Algeria, Greece, Italy, Morocco, Portugal, Spain, Tunisia, Turkey, and the former Yugoslavia (Van Mol & De Valk, 2016). Among the countries included in this study, France, Germany, and the Netherlands were the main destinations for these migrants. Simultaneously, decolonisation brought about large migration flows from the colonies towards the former colonial power countries, including France, the Netherlands, the UK, and in the 1970s, Portugal (Van Mol & De Valk, 2016). In addition to workers and people from former European colonies, at the end of the last century, asylum seekers became another important category of migrants from non-European counties. Especially in the 1980s and after the fall of the Berlin Wall, the numbers of asylum applications started to rise in Europe (Hansen, 2003). To date, Germany has been one of the largest recipients of asylum applications in Europe, but France, the Netherlands and the UK have also received significant numbers of asylum seekers among our study countries.

By contrast, Italy, Portugal, and Spain had long been emigration countries. However, the restrictions on the entrance of foreigners into North-Western Europe diverted significant migrant flows towards Southern Europe, especially in the 1990s. This was the period of immigration legislation and entrance control systems in the North, alongside slow economic growth and falling birth rates which in turn resulted in labour shortages in the South. Therefore, Southern Europe became an attractive destination for non-European migrants, especially those from North Africa, Latin America, Asia (Castles, 1998) a trend not repeated in the Western European countries in our study (Germany, France, UK, and the Netherlands). While post WW2 migration to this part of Europe was focused mainly on migrant workers taking up low-paid jobs, immigration of students and highly skilled workers has become increasingly important since the 1990s. In this context, several European countries, including France, Germany, the Netherlands, and the UK simplified procedures for international students to make the transition to the labour market in the destination country (Tremblay, 2005; Van Mol, 2014).

Although we can identify some similarities in groups of countries (for example, those in Southern Europe are more similar to each other than they are to those in North-Western Europe), given the quite different ethnic structures of the countries in this study, we can expect that also the spatial patterns of ethnicity differ. However, to fully understand segregation, we must introduce the notions of various dimensions and spatial scales.

## 2.2 Dimensions of residential segregation

Residential segregation is the phenomenon of different population groups living in different parts of the same (urban) space. However, living apart can be considered and measured from different perspectives. Searching for the most appropriate indices to study segregation in the US, Massey and Denton (1988) identified five distinct dimensions of segregation: centralisation, evenness, clustering, exposure, and concentration. In summarising these dimensions, Massey and Denton also established suitable indices for measuring each of them.

Each dimension refers to a specific aspect of segregation, starting from the fact that to achieve segregation a minority group usually needs to be unevenly distributed across the city – in other words, have a greater presence in some neighbourhoods than in others deviating from the city on average (evenness dimension). (Un)even distribution of ethnic groups is important for the social interactions in the city: if the minority population mainly lives with members of the same group in their neighbourhood, this limits their potential interaction with other groups (exposure dimension). This is important from an individual perspective, for example for the integration of migrants in the receiving society. For a long time, large urban areas have been viewed as having a core and a periphery. The centralisation dimension measures the residential over-representation and under-representation of population groups in the urban core. By being over-represented in cores or peripheries, minority groups will have different access to housing, education, work, healthcare, etc. Minority groups can also be scattered around the urban area, but they may also settle in a way that they form large contiguous enclaves (clustering dimension). Such clustering of minorities may be representing housing patterns of the city, as well as the distribution of ethnic cultural facilities (churches), ethnic shops etc. Minority groups may be concentrated within densely built-up but geographically small areas, occupying less physical space than the majority group (concentration dimension). So, although we start from the observation that segregation implies living apart, the geographies of segregation are more complex, and each dimension captures a single aspect of this complexity.

While exposure deals with the question by whom people are surrounded in their own neighbourhood, evenness, clustering, centralisation, and concentration all look at how different groups of people are geographically distributed across urban space. Reardon and O'Sullivan (2004) argued that it was not necessary to have evenness and clustering as separate dimensions – they reworked and collapsed them into one, as two opposite sides of the same dimension. Clustering takes into account the spatial arrangement of units, in other words that people who are closer to each other are more similar than the ones who are more distant and as such it puts more weight on closer neighbours. So clustering sheds even more light on uneven distributions, resulting in more clear clusters – areas with similar people.

Evenness in its original form, as defined in Massey and Denton (1988), is not a spatial dimension, because the same index value can be obtained with substantially different arrangements of neighbourhoods. This holds for discrete non-overlapping spatial units, but in the last decade, overlapping spatial units at multiple scales have been increasingly used (Hipp & Boessen, 2013; Östh, Clark, et al., 2014; Petrović et al., 2018). Starting from the central spatial unit, neighbouring units are added at increasingly large spatial scales, forming the so-called bespoke multiscale areas (Petrović et al., 2018). Using the approach of bespoke multiscale neighbourhoods, every measure of segregation becomes spatial. This paper therefore keeps the original dimensions, including evenness and clustering, as suggested by Massey and Denton (1988). This way we explicitly take into account an important issue of spatial scale as well as the spatial arrangement of units in measuring each dimension of segregation.

## 2.3 Spatial scale, urban form and residential segregation

Studies of residential segregation centre around measuring individuals (or collections of individuals) in areas, often identified as neighbourhoods. Conceptually, this is challenging, because a neighbourhood can mean many things to different people and, the administrative units in which population are routinely reported rarely conform to all (or even any) of the neighbourhoods that people on the ground experience (see for instance Talen, 2018). However, the issue is more complex than simply one of conceptual deviance from notions of neighbourhood to the measured reality: each urban area has its own size and its own urban form. The size of neighbourhoods may be linked to the size of the urban space, such that larger urban areas correspond with larger neighbourhoods. Conversely, smaller, compact cities may also have smaller compact neighbourhoods. And of course, there is no reason to assume that the scale of neighbourhood remains constant within a single urban space and as a result, it could vary within cities as well. Crucially, various spatial contexts are important for individuals, starting from their immediate neighbourhoods, facilitating social contacts, up to regional labour markets (Petrović et al., 2020). As such, spatial scale is one of the key challenges when measuring segregation (see, for eaxmple, Reardon et al., 2008; Manley et al., 2015). With the exception of a handful of studies, (see, Östh, Clark, et al., 2014; Manley et al., 2015; Marcińczak et al. 2021), most studies adopt a single (administrative) scale.

We take a different approach, which does not consider spatial scale to be a problem, but a means through which is it possible to explore how different processes occur in neighbourhoods of varying extents across cities (Petrović et al., 2020). In this study, spatial scale is a means of creating 'distance profiles' of segregation, conceptually similar to segregation profiles introduced by Reardon et al. (2008), making it possible to look at the urban environment in a continuous way and to observe how the extent of segregation gradually changes (or remains constant) with changing scale. Ultimately, by doing this we enrich the measures of segregation with the important element of spatial scale.

Furthermore, fixed spatial units are limited by the city boundary, while people who live close to the city boundary may still be exposed to the surrounding areas. A surface-based approach of segregation by O'Sullivan and Wong (2007), demonstrated the importance of looking at urban space continuously beyond the city boundary when studying segregation, although their study used only one spatial scale. Generally, studies which compare different scales find that segregation decreases with spatial scale (Reardon et al., 2009; Östh, Clark, et al., 2014). This can be related to averaging out the population counts at larger scales, but also to using fixed spatial units limited by the city boundary. Taking into account the periphery of the city would offer complementary information about segregation, especially how it changes across various scales of urban space.

## 3 Data and methods

We study five dimensions of ethnic segregation, measured at a range of spatial scales, for seven European capital cities. We use data provided by the European Commission from the D4I (Data for Integration) data challenge, sourced from the national statistics offices in the Netherlands, Germany, Portugal, UK, Spain, France, and Italy (see Alessandrini et al., 2017). The dataset contains harmonised, high resolution spatial data reported in regular grid cells (100m by 100m) recording the ethnic origin of migrants. Categorising immigrants is challenging, since each country has its own immigration history, mix of ethnic groups and definitions to group ethnicities. In defining ethnic groups, some countries rely on citizenship, while others rely on the country of origin, but for comparative analysis, a single definition must be used. Large European cities are characterised by ethnically mixed neighbourhoods, within which several different ethnic minorities live together. Andersen (2019) therefore argues that in European countries, ethnic segregation should be studied between the native population and ethnic minorities in general, particularly those of non-Western origin. For this study we adopt the Non-Western definition used by Statistics Netherlands (see Alders, 2001), as it is one of the most straightforward definitions which can be applied across all countries. To do so, the native population together with Western immigrants are compared to non-Western immigrants, where non-Western immigrants originate from African, Asian or Latin-American countries[1].

---

[1] The data from Portugal does not distinguish people by the country of origin. Therefore, non-Western migrants in Lisbon also include people from Anglo-America, because they cannot be separated from Latin-American migrants.

To define urban areas we used Functional Urban Areas (FUA) developed by the OECD and the EU. This definition increases the comparability of the economic, social and environmental performance of metropolitan areas (OECD, 2012). The FUA consists of the densely populated core and hinterland (periphery), with a labour market which is highly integrated. We have applied the core and hinterland boundaries on the grid cells provided in the D4I data, and this consistency in the definitions of the FUAs allows us to compare the urban spaces in the seven countries.

We examined five dimensions of segregation (an overview of the dimensions, corresponding indices and formulas is available as Supplementary Material). The first is *centralisation*, which directly uses the FUA definitions to measure the relative concentration of the two groups in the urban core. It ranges from 0 to 1, and represents the proportion of members of one group (Western or non-Western) living in the urban core. The second dimension, *evenness*, is quantified using the index of dissimilarity, and reported as the percentage of the members of a group who would need to move in order to achieve an even distribution of that group across the whole city. For example, a value of 0.4 means that 40% of the minority population group would need to move to achieve an even distribution. The index ranges from 0, when the share of migrants in the neighbourhood is the same as in the entire urban area and there is no segregation, up to a value of 1, when there is complete segregation.

While evenness compares neighbourhoods' ethnic composition with that of the average of the FUA, *exposure* explicitly takes into account the relative size of the non-Western population: if the share of non-Western population is high in the city, then they are less likely to be exposed to the other ethnic groups in their neighbourhood of residence. We use the isolation index to quantify this experience. Where a low value is reported, spatial isolation is low and a high potential for inter-ethnic interaction exists, and vice versa. *Concentration* compares the size of the population and the area where this population is located. The index is high when non-Western people are located in a smaller number of 100m by 100m cells (i.e. a smaller part of the urban area). It simply shows the percentage of non-Western people that would have to shift spatial units to achieve a uniform density of this group over all areas. For example, a value of 0.4 means that 40% of non-Western people would need to shift units to have a uniform density of this group over all units.

The final dimension is *clustering*, measured by the spatial proximity index, which shows to which extent spatial units with many non-Western people locate close to each other. This index equals 1 if there is no difference in clustering of Western and non-Western people, and is greater than 1 when members of each group live closer to the members of their own group than to the ones from the other group. The index below 1 is unusual, because it represents exceptional cases when members of two groups reside closer to each other than to members of their own group. Clustering is therefore 'the most spatial' dimension we use, because it also includes the distance decay, putting more value to the near than to the distant spatial units.

All indices depend on the size of neighbourhoods used in urban space. Standard administrative units, although practical, do not represent social processes such as segregation; they have different sizes in different countries, are inconsistent over time, and individuals are not always centrally located within them. Individuals who live close to the edge of unit may be more connected with people in a neighbouring area than in their own unit. Moreover, small spatial scales may be insufficient to represent an individual's local environment and so it is important to use neighbourhoods of multiple sizes to better characterise the potential experiences individuals may have (Petrović et al., 2018). To examine how segregation continuously changes, we measured it over 101 different spatial scales of bespoke areas. We started from small spatial building blocks of 100m by 100m grid cells, and increased the radius in 100m increments up to a 10km radius, which required much computational power and time. The smaller areas capture the immediate surroundings of individuals while the bigger areas, with greater overlap delineate increasingly wide city contexts. Using these areas, we compute distance profiles for each of the dimensions of segregation of non-Western migrants in all seven capitals. The distance profiles show how segregation changes as an individuals' neighbourhood border moves further away from their home: it starts from the smallest residential context and expands towards larger areas (see Petrović et al., 2018).

Before proceeding to the results, two data issues should be highlighted. First, ethnicity is defined in different ways in different countries (birth, citizenship, or a combination of both). This, however, is not a problem, because it reflects how ethnic minorities are defined in certain contexts. For our analysis, we do not need the same definition of ethnicity, but rather definitions that represent what 'ethnic minority' means in each country. Second, the total population counts were estimated by the data provider for each 100m by 100m cell, using census data combined with land use data and building footprints. Consequently, these numbers are close to the observed numbers but have a degree of uncertainty around them. By contrast, the shares of ethnic groups in the total population come from the census areas (postcode areas in some European countries). This may lead to under- or overestimation of the shares of ethnic minorities at the very smallest scales. However, the data still represents the most advanced collection for such a large number of countries.

## 4. Results

We start by introducing the cities with their cores and hinterlands, by mapping the distribution of ethnic minorities within them. We then develop the analysis by reporting the first dimension of segregation – centralisation – to explore the core and hinterlands. Finally, the other dimensions, which are more complex and multiscale, are presented by first focussing on micro-segregation and finally moving to the multiscale segregation.

Figure 1 reports the maps showing the share of non-Western migrants in grid cells in our case study cities. London has the most striking areas with high shares of non-Western migrants: many small neighbourhoods with a majority of non-Western people cluster in large areas of the urban core. Non-Western migrants concentrate in urban cores in most of the other cities, especially in Amsterdam and Berlin, perhaps because urban cores of these cities offer more economic and housing opportunities for migrants, and because some neighbourhoods function as entry gates for migrants. In the South European capitals (Lisbon, Madrid, Rome), non-Western migrants are more equally spread over the core and hinterland than in the Northern European capitals. In Madrid, where the majority of non-Western migrants have Latin American origins, there is a clear hinterland pattern complementing the core. This pattern is also visible for the relatively small group of African migrants, potentially because of a greater accessibility and affordability of housing in the hinterland. However, in Madrid and Lisbon language and job opportunities may also determine the distribution of non-Western migrants. Those Latin

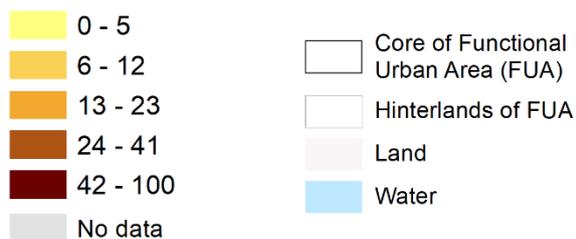
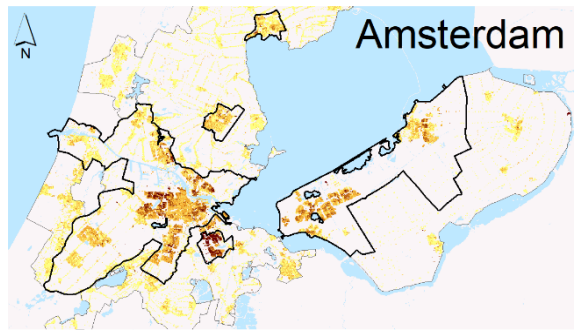
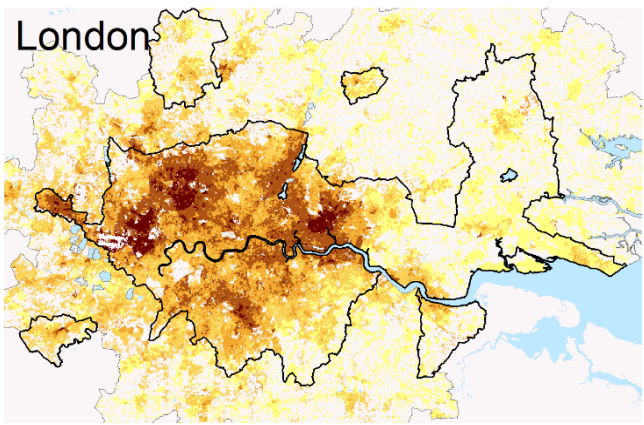
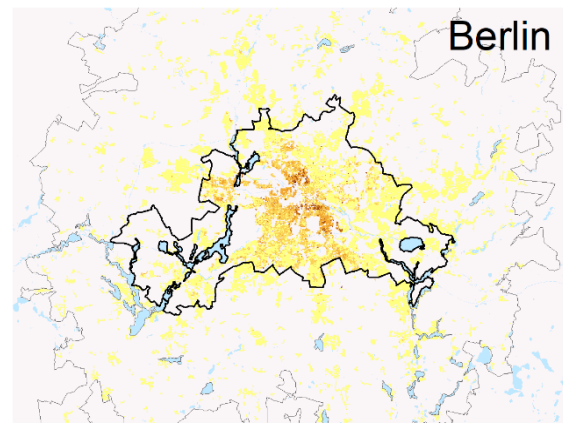
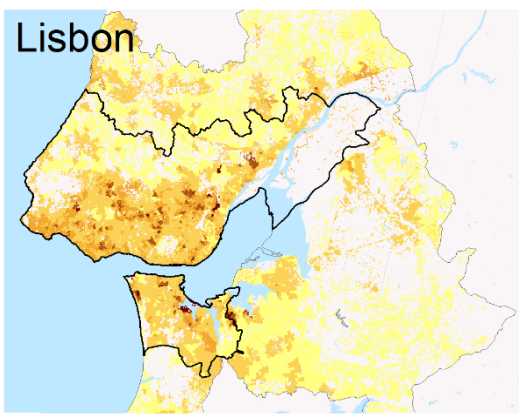
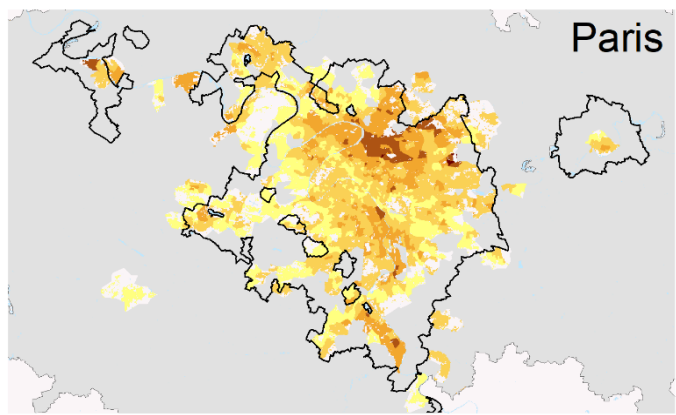
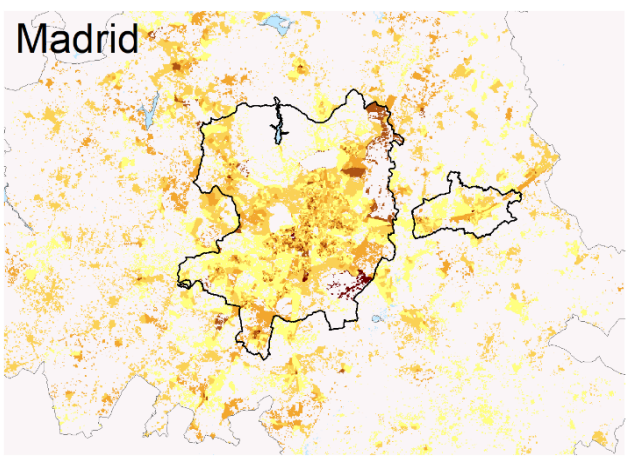
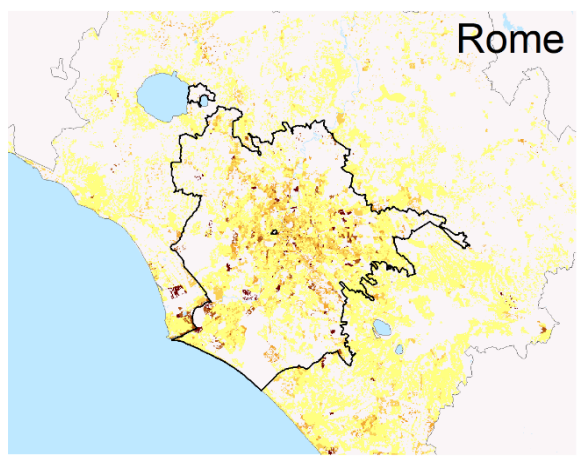
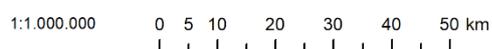

Figure 1: Share of non-Western people in 100m by 100m grid cells in metropolitan areas (cores and hinterlands) of seven European capitals

American immigrants who share their language with the native population, can find jobs in the private, and public sectors inside and outside the urban core. Due to these greater opportunities, they are scattered over the FUA of Lisbon as in Madrid. Lisbon also has the largest African population percentage-wise among all the seven cities, and this minority group tends to concentrate in the core city area. This leads to a different spatial pattern compared to Madrid. The insights from the maps regarding migrants' locations in urban cores as opposed to hinterlands bring us to the first dimension of segregation which we examined: centralisation.

## 4.1 Centralisation

Figure 2 reports the proportion of non-Western people living in the densely populated and economically stronger urban cores of the seven FUAs. Using the centrality index, we can see that in Berlin the non-Western migrants almost exclusively settle in the urban core, as highlighted on the Berlin map. The centrality index also confirms the insights from the maps that non-Western people are the least centralised in Madrid. The biggest difference between the indices for the two groups occurs in Amsterdam, with non-Western people tending to locate in the urban core more than Western individuals, a pattern also repeated in Rome. Madrid and Rome are, therefore, the metropolitan areas whose hinterlands have received considerable shares of non-Western people in that setting. However, the share of the population in these group is very different in these two cities (Figure 3). Overall share of non-Western people in Rome is low in comparison with Madrid which has a considerable share of non-Western people. By comparison, London is the city with the highest share of non-Western people, who are very centralised in the urban core.

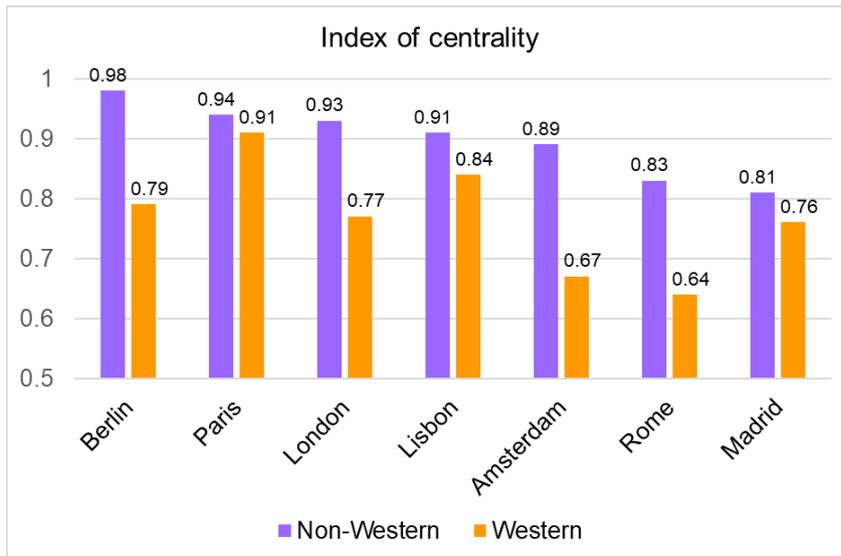

Figure 2: Index of centrality of Western and non-Western people

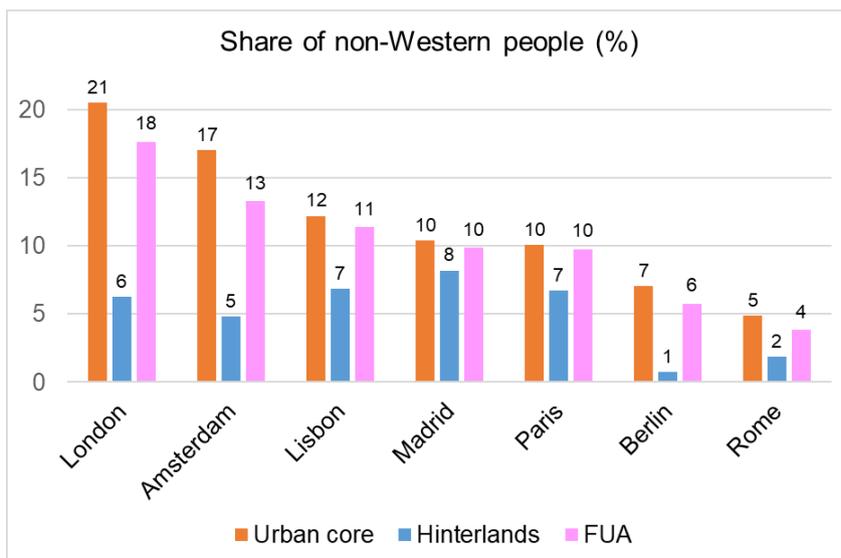

Figure 3: Share non-Western people in different parts of FUA

## 4.2 Micro-scale segregation: The analysis of the lowest spatial scale

So far, we have compared the urban cores and hinterlands, in terms of the presence of migrants. However, migrants can be unevenly distributed within both of these areas and so it is important to examine the other dimensions of segregation. Together, all five dimensions allow comparison of different aspects of segregation. Figure 4 presents the remaining 4 dimensions at the lowest available spatial scale (100m by 100m grids) by the seven capitals, distinguishing FUA cores and hinterlands. Clustering has similar, high, values in almost all cities because, by default, it is measured at a different scale – it has values around 1 or higher (rarely lower), while centralisation, evenness, exposure, and concentration may take values between 0 and 1. London has the highest clustering of all the cities and in combination with the high isolation index, London is a city in which the non-Western population are least exposed to Western people. By contrast, in Berlin non-Western people are more exposed to Western people, although their distribution remains uneven and they are still concentrated in small neighbourhoods.

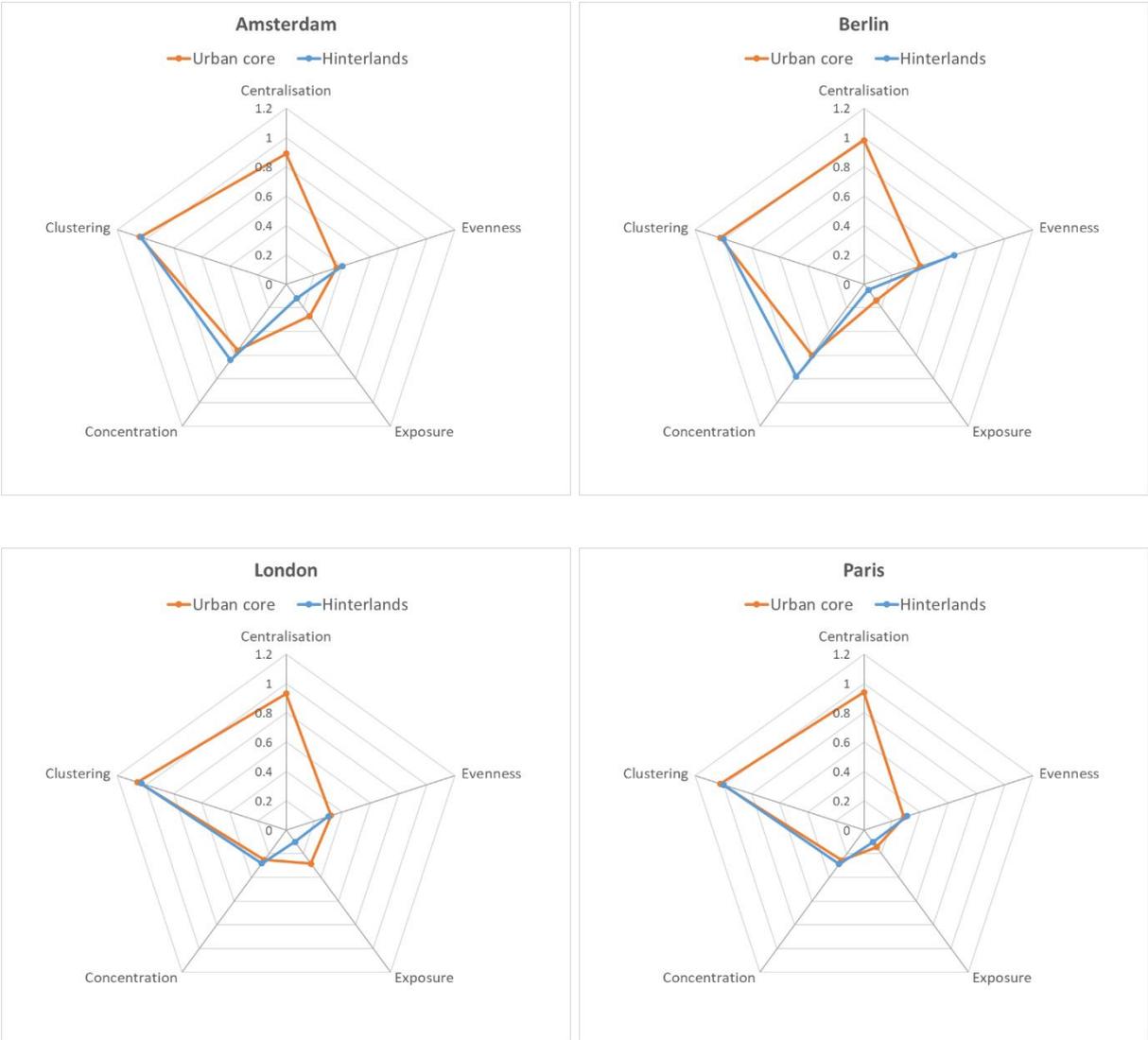

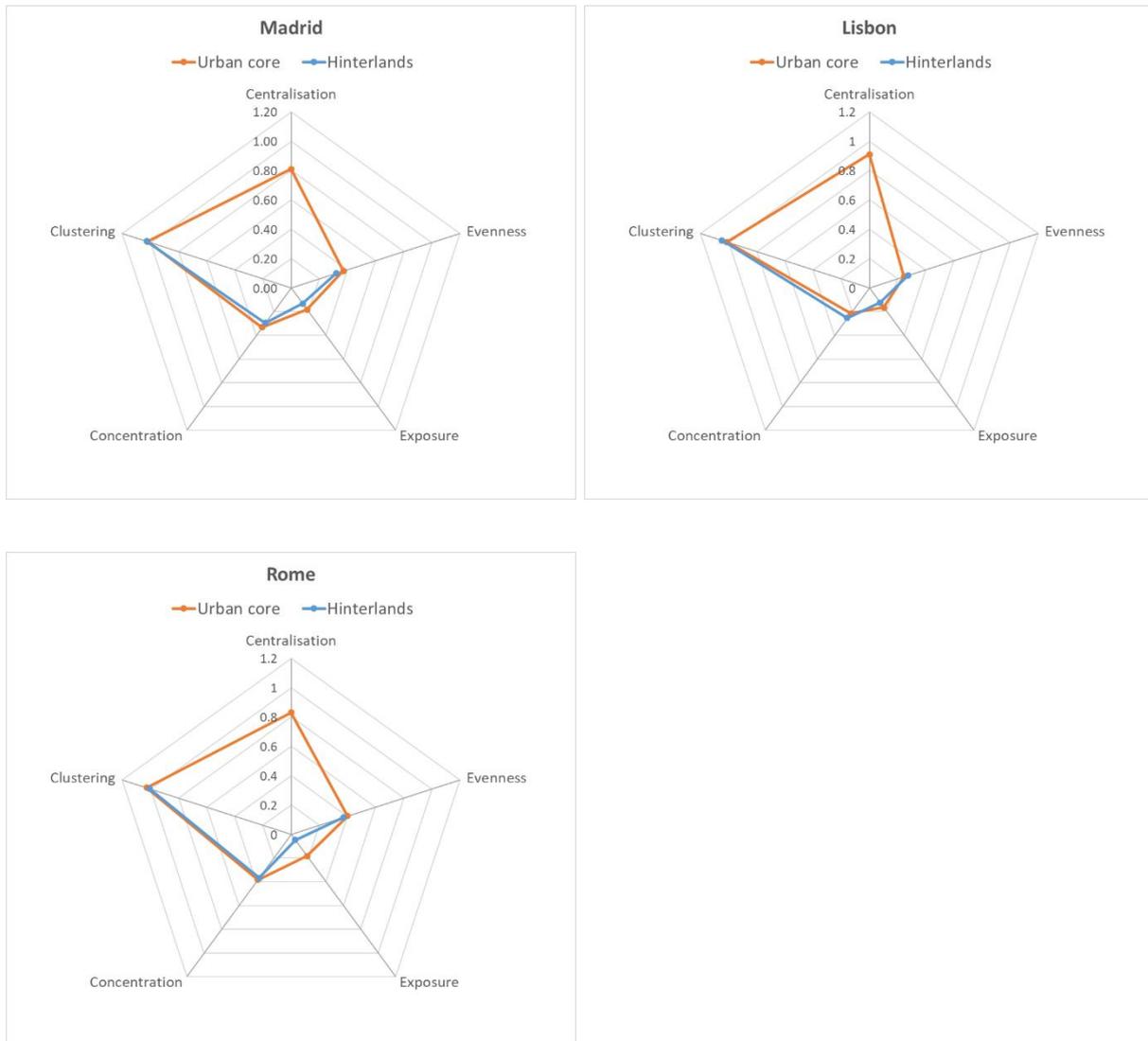

Figure 4: Five dimensions of segregation measured at the smallest spatial scale (100m by 100m) for FUA cores and hinterlands (centralisation is by definition measured only for the urban core)

In addition to different dimensions, we also compare the city cores and hinterlands. In Madrid and to some extent in Lisbon small neighbourhoods in the city core have similar segregation values as those in the hinterland, this is not replicated in the other cities. The differences between urban cores and hinterlands also vary across the dimension: non-Western people are residentially more isolated (that is, less exposed to Western people) in the city cores, while they are more concentrated in the hinterlands. This is related to the fact that at the point of arrival, migrants usually go to the city cores and they experience greater exposure to their own group members than to the native population. In the hinterlands they are concentrated in specific neighbourhoods.

## 4.3 Multiscale segregation

In addition to the above analyses for smallest spatial scale (100m by 100m), we can use residential areas of various sizes to measure evenness, exposure, concentration, and clustering. The size of neighbourhood matters: An individual can have no non-Western migrants in the immediate surrounding of their home, but as the spatial reach of their neighbourhood extends, they can have many more non-Western neighbours. Although with increasing scale the distance from home increases, it is still possible that people meet, but instead of meeting in the street, the interactions might happen at neighbourhood cafés or via children in the local school. In this case, non-Western migrants are underrepresented in the immediate neighbourhood, and overrepresented in the bigger areas. To explore this, we measured the evenness, exposure, concentration, and clustering at the range of 101 spatial scales and computing distance profiles.

### 4.3.1 Evenness

Figure 4 depicts the distance from home on the horizontal axis, and the dissimilarity index value on the vertical axis (higher values show greater unevenness). By plotting the capital cities together, we can compare how segregation changes with scale and by country. In most of the cities, non-Western migrants are more unevenly distributed in the hinterland than in the urban core: Hinterlands generally have fewer non-Western people than the urban cores, but these people locate in specific parts of the hinterland, most likely in places where they can access and afford housing or settle close to family. This occurs at different spatial scales depending on the city. In Berlin's hinterland, with a low percentage of non-Western migrants, the minorities are particularly overrepresented in neighbourhoods under 1km radius. By contrast, in Amsterdam, whose hinterland has larger share of non-Western people than Berlin, the overrepresentation occurs for scales up to 5km radius. The distance profile of the Amsterdam's hinterland highlights that areas with high shares of non-Western people are located close to areas with lower shares, forming a mosaic structure. However, within 3–5km, the share of non-Western migrants grows, reducing the variation in exposures. Using the distance profiles of segregation, we can identify the spatial scales with high segregation in each city. While unevenness often decreases with the increasing scale, the 3.5km scale in Amsterdam hinterlands demonstrates that this is not always the case.

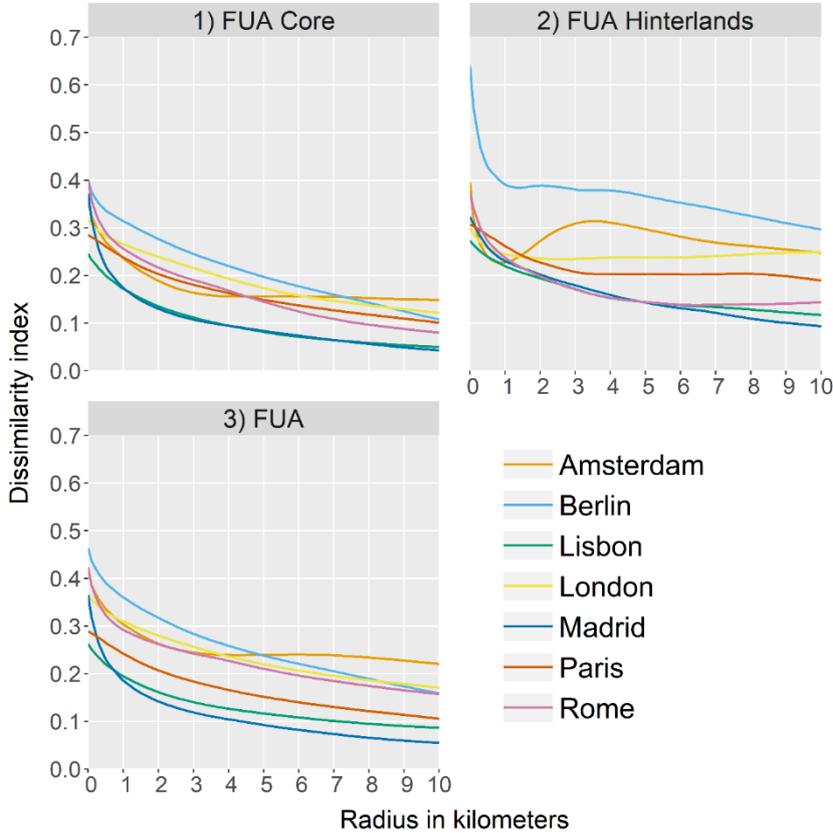

Figure 5: Distance profiles of the dissimilarity index

The results for the range of scales suggest that measuring evenness at single administrative scales obscures smaller-scale neighbourhood ethnic compositions. For example, although at larger scales almost evenly distributed, non-Western migrants in Madrid remain segregated in smaller neighbourhoods, particularly in the urban core. Although Lisbon and Madrid have similar low levels of segregation at the larger scales, the fine-grained spatial scale reveals substantial difference, with greater local segregation in Madrid than Lisbon. By contrast, Madrid is the *least* segregated city at the largest scale but, with Rome and Berlin, one of the *most* segregated city at the smallest spatial scales.

### 4.3.2 Exposure

Exposure dimension of residential segregation hinges also on the city-wide proportion of the minorities. Hence, we are particularly interested in the scale effects when it comes to exposure. Rome, with the lowest overall city-wide share of non-Western people, also has the lowest residential isolation and highest potential of meeting and interaction between our two ethnic groups (see Figure 6). Despite this low isolation index value at most spatial scales, the isolation of the non-Western group in Rome at the smallest spatial scale, is larger than in Berlin, Madrid, Paris and Lisbon. Thus, the immediate environment outside the front door is more segregated in Rome. The isolation index reveals a different picture in London when compared with the results for evenness, which is related to the higher city-wide share of non-Western people. Hence, the isolation index of the non-Western group, particularly in the urban core, is distinctly higher than in other cities.

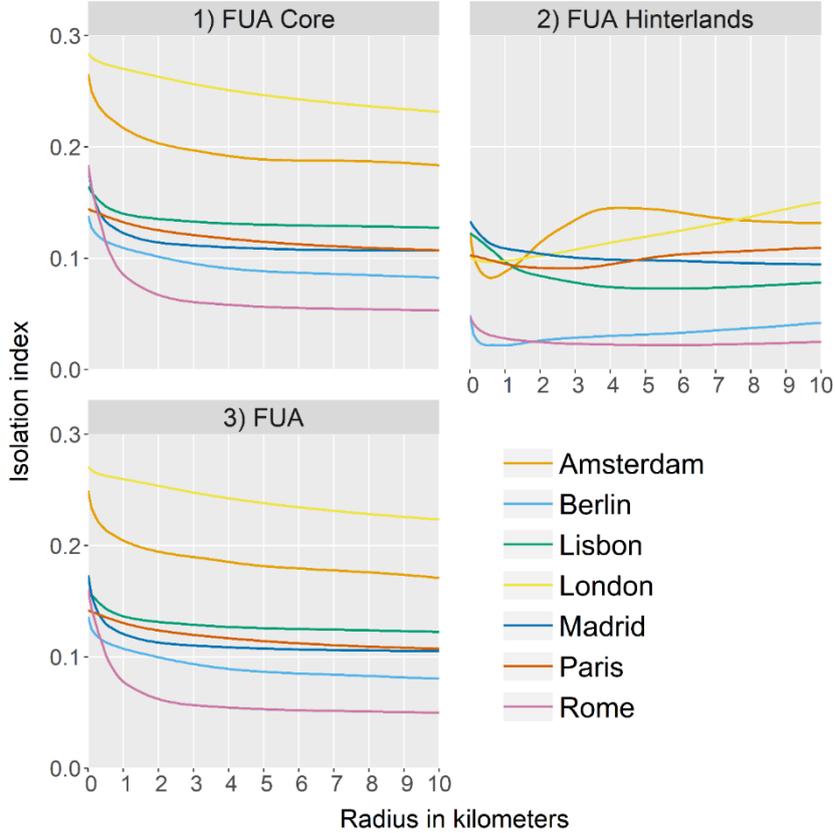

Figure 6: Distance profiles of the isolation index

Although persistent segregation over multiple spatial scales is a common phenomenon in Europe, some cities, such as Madrid or Amsterdam, have greater variation in segregation than others. Furthermore, the indices of Rome or Berlin, which have different levels of segregation (comparing to other cities) in different dimensions (low isolation, but high dissimilarity), show that spatial patterning of segregation is very different and thus needs to be measured both along multiple scales and multiple dimensions.

### 4.3.3 Concentration

All seven cities have quite high concentration values, meaning that non-Western people live in a small number of spatial units rather than being scattered around the urban area. This is particularly true for the scales up to 2-3km. There is, however, variation across spatial scales, depending on the city. The best examples are Rome, Madrid and Lisbon, all with the same concentration values at the smallest scales – up to 1km – in the city cores. Beyond 1km, Rome has considerably higher concentration values.

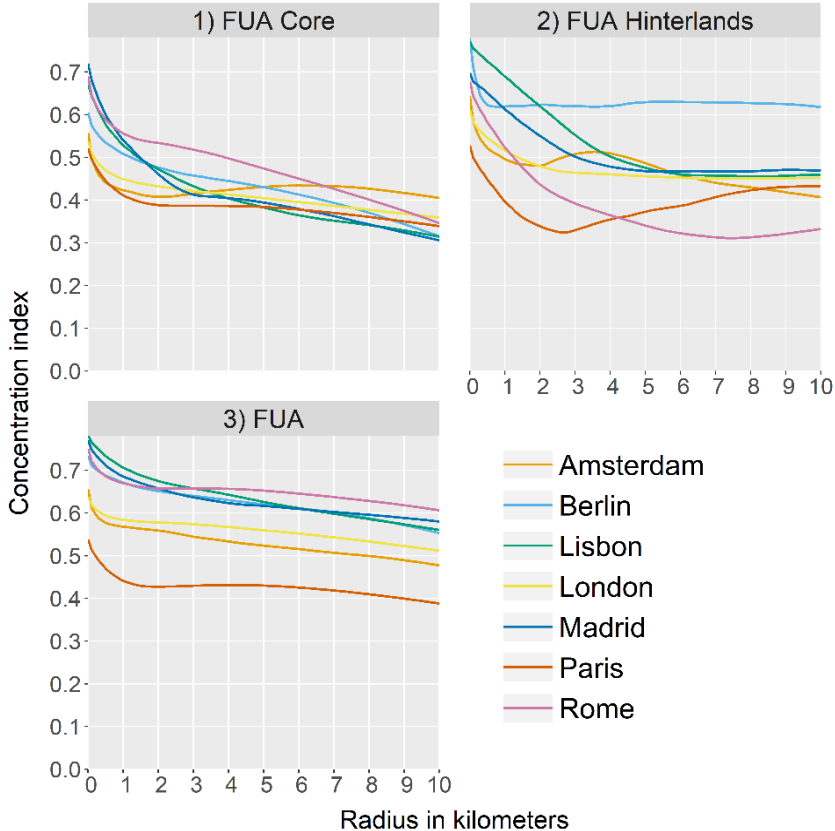

Figure 7: Distance profiles of the concentration index

In comparison to other dimensions, London and Paris now do not score very high. Although segregated according to the isolation index, the concentration of residential areas of non-Western minorities are scattered around the city. Amsterdam shows the unusual outcome of increasing segregation with spatial scale, both in the city core (around 7km) and especially in the hinterland (around 3-4km). Similarly, Paris shows an increasing segregation at larger scales as well, but we need to be cautious here, because this may be partially due to the many missing data in the Paris hinterland. In any case, concentration and other dimensions of segregation can increase with spatial scale when we consider space as continuous, using bespoke overlapping spatial units and look separately at city cores and hinterlands.

**4.3.4 Clustering**

While the previous dimension measured to which extent the non-Western people concentrate in certain neighbourhoods, clustering measures to which extent the neighbourhood with many non-Western people tend to group and co-locate in residential space. Figure 7 reports the spatial proximity index. The peaks denote the spatial scales at which clusters of non-Western minorities are formed. This is the most notable in Amsterdam hinterlands at the scale of 3.5km. We can therefore clearly recognise areas with a 3.5km radius where non-Western people cluster in the peripheral parts of Amsterdam. On the contrary, clusters of non-Western minorities in Amsterdam city core are of much smaller size – they mostly form at the scale of a few hundred metres, with a gradually decreasing scale. Besides Amsterdam, London is the most prominent city when it comes to clustering of non-Western minorities in the core. This city has the high clustering value at most of the spatial scales in urban core, including larger ones. So, although the index is high for smaller scales, these smaller clusters group to form bigger ones.

In the other cities, clustering varies both across spatial scale and between the city cores and hinterlands. In most of the city cores, there are smaller clusters, while at the other scales the index remains close to 1, that is no differential clustering between the Western and non-Western groups. In the hinterlands, the cities become more differentiated, particularly when comparing small-scale clusters in the hinterlands and city cores: Lisbon and Madrid have stronger clustering of non-Western people at smaller scales in the hinterlands than in the city cores. By contrast Rome and Berlin have even lower clustering values at the smallest scales in the hinterland than in the city cores. We also see unusual values of the clustering index below 1 in Berlin, meaning that non-Western people are closer to the Western people than to their own group members. This is the result of a large population of Western people, combined with a non-Western population scattered around the city, particularly in the hinterland (see Figure 1).

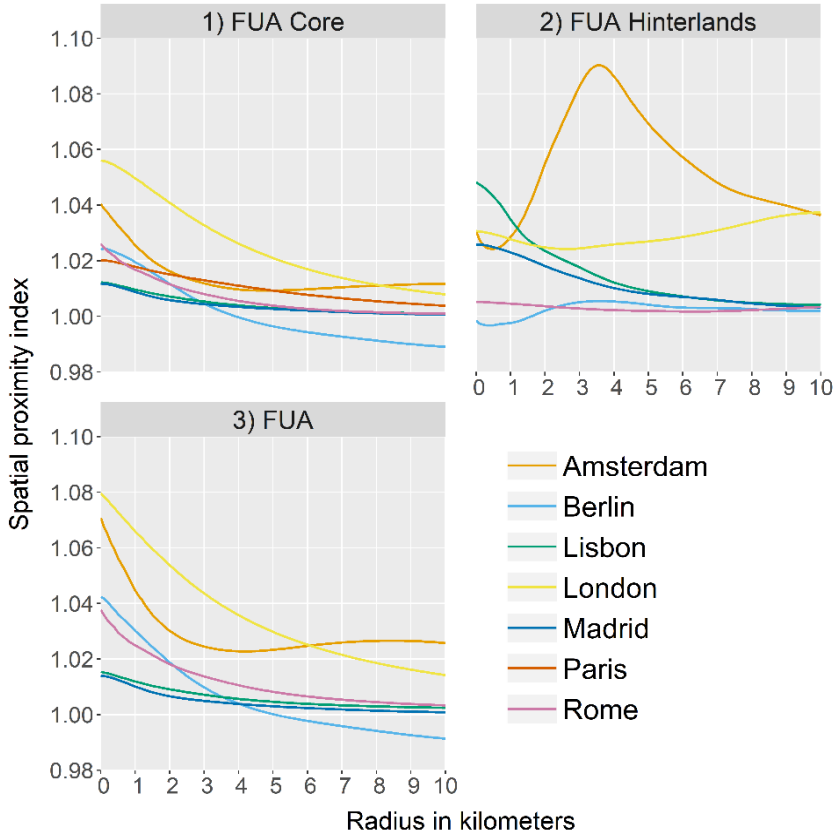

Figure 8: Distance profiles of the spatial proximity index

# Discussion and conclusions

In this study, we examined segregation in seven urban regions in Europe from a bespoke, multiscale and multidimensional perspective. We used five dimensions of segregation to understand the potential for people to meet, which is crucial for the functioning of cities as integral urban systems. Overall, we identified differing levels of residential segregation and potential exposure, and we found that neighbourhood size and local context matter in terms of the share of non-Western migrants and how ethnic groups potentially experience segregation. This gave insight into segregation at a range of spatial scales for multiple dimensions and multiple cities. While our study confirms that South-European capitals have lower levels of ethnic segregation than Western European capitals, different dimensions and scales shed more light on segregation across Europe.

The harmonised definitions of FUAs used in the study facilitate international comparability of the economic, social and environmental experiences of metropolitan areas. One of the social aspects of a city is how it deals with segregation and the socio-spatial integration of migrants. Using the FUAs, we compared segregation in densely populated urban cores and in the related hinterlands, and the different dimensions we examined tell us about different aspects of segregation: Evenness showed that in Berlin, Rome and Madrid evenness is experienced very different in smaller neighbourhoods than for the city has a whole. By contrast, exposure demonstrated that spatial isolation of large non-Western populations was greatest in London, then Amsterdam, and therefore the potential for interaction with the Western population in residential neighbourhoods is low. This is in line with the previous knowledge about high segregation in London (Johnston et al., 2016). However, our study provided many new insights on residential segregation in London. Concentration of non-Western people was, on the contrary to exposure, low in London, due to the large non-Western population residing in large areas of the city, which was sometimes clustered, as the last dimension showed. The defining features of London ethnic segregation, compared to the other cities, are a large non-Western population concentrated in large areas, forming substantial clusters, but also resulting in the low meeting potential between non-Western and Western groups (measured by exposure), persistent at various spatial scales. Finally, Amsterdam demonstrated most clearly how segregation (for multiple dimensions) increases with the spatial scale, which is in contrast with most of the previous findings in the segregation literature.

Berlin, compared to London and Amsterdam, has a smaller, more unevenly distributed non-Western population, with smaller clusters. Paris is more similar to London and Amsterdam in terms of the low concentration of non-Western migrants, but it also has lower isolation, with greater potential for meeting between groups. However, due to the missing data in the outer ring of the urban core and most of the hinterlands, segregation in Paris is most likely underestimated. Still, big clusters of minority concentration can be found in the hinterland of Paris that is very different from the other urban areas.

While the low meeting potential between the non-Western and Western groups prominently concerns urban cores, in Madrid and Lisbon, urban hinterlands do not have lower levels of segregation than the urban cores, in line with previous literature regarding the peripheralisation of ethnic minorities in Southern European cities (Arbaci, 2008).

The South-European cities (Madrid, Lisbon and Rome) are furthermore particularly interesting when it comes to variations in segregation across spatial scales. With relatively small (Madrid and Rome) and medium (Lisbon) non-Western populations, concentrated in smaller residential areas as found also in previous research (Leal 2016), these cities should particularly focus on small spatial scales to combat segregation. The segregation drops when non-Western people move from their immediate neighbourhood, so they are not clustered at higher scales, i.e. they live in smaller neighbourhoods scattered around the city. Higher segregation levels and more persistent segregation at increasing scales, such as in the case of London, require more complex actions implemented multiple spatial scales.

Examining the segregation of ethnic groups at various spatial scales allows us to design policy interventions at the most efficient spatial scales. Comparative analysis of multiple countries provides a context within to understand the differences between cities in different countries. However, different concepts of ethnic origin, such as place of birth and citizenship, in different countries need to be considered. Residential segregation is related to segregation in other life domains such as education, work and free time (van Ham & Tammaru, 2016; Tammaru et al. 2021). Spatial scale and the dimension of residential segregation affect segregation in other domains. For example, the scale of residential segregation suggests whether schools in specific parts of the metropolitan areas are also segregated or how big are potentials for inter-group interaction in public spaces. In light of the recent immigration to Europe, the analyses should be further developed by focussing on subgroups of the non-Western migrants, such as African, Asian and Latin-American. These groups have different residential preferences as well as economic and cultural requirements, which are likely to result in spatial patterns of segregation along multiple scales and multiple dimensions.

To conclude, this study showed in much detail that we should be careful when interpreting a single measure of residential segregation at a single spatial scale. The reality is that segregation measures vary considerably over scales and different dimensions. This is not just a methodological point however: there are substantial and important empirical differences which are related not only to the composition and diversity of the groups under investigation – something we kept constant between cities by using a single definition for non-Western minorities – but also that the differences in how segregation plays out is a function of the urban form of the urban space. Each city has its own perspective of the urban space, which includes various scales of residential environments. When comparing cities, we should be careful when using single dimension and scale, because this may lead us to incomplete conclusions. Finally, various dimensions and scales yielding different results are not conflicting – they provide complementary information on segregation levels in urban space.

# Supplementary material

| Dimension | Index | Formula |
|---|---|---|
| Centralisation | Centrality, using urban core and the whole FUA | $PUC_x = X_{UC}/X_{FUA}$ <br> $PUC_y = Y_{UC}/Y_{FUA}$ |
| Evenness | Dissimilarity between non-Western and Western groups | $D_s = \frac{1}{2}\sum_{i=1}^{n}\left\lvert\frac{x_{si}}{X_s} - \frac{y_{si}}{Y_s}\right\rvert$ |
| Exposure | Isolation of the non-Western group | $_xP_{xs} = \sum_{i=1}^{n}\left(\frac{x_{si}}{X_s}\right)*\left(\frac{x_{si}}{t_{si}}\right)$ |
| Concentration | Concentration of the non-Western group | $DEL_s = \frac{1}{2}\sum_{i=1}^{n}\left\lvert\frac{x_{si}}{X_s} - \frac{a_{si}}{A_s}\right\rvert$ |
| Clustering | Spatial proximity between non-Western and Western groups | $P_{sxx} = \sum_{i=1}^{n}\sum_{j=1}^{n}\frac{x_{si}x_{sj}c_{ij}}{X_s^2}$ <br> $P_{sxy} = \sum_{i=1}^{n}\sum_{j=1}^{n}\frac{x_{si}y_{sj}c_{ij}}{X_sY_s}$ <br> $c_{ij} = \exp(-d_{ij})$ <br> $SP_s = (X_sP_{sxx} + Y_sP_{syy})/T_sP_{stt}$ |

$PUC_x$ – proportion of non-Western people living in the urban core

$X_{UC}$ – number of non-Western people in the urban core

$X_{FUA}$ – number of non-Western people in the FUA

$PUC_y$ – proportion of Western people living in the urban core

$Y_{UC}$ – number of Western people in the urban core

$Y_{FUA}$ – number of Western people in the FUA

$D_s$ – dissimilarity index at scale $s$

$s$ – scale of bespoke area; $s = 0, 100, 200, ... , 10000m$

$n$ – number of grid cells the urban area (core, hinterlands, FUA)

$x_{si}$ – number of non-Western people measured at scale *s* for grid cell *i*

$X_s$ – number of non-Western people measured at scale *s* for the whole urban area (core, hinterlands, FUA)

$y_{si}$ – number of Western people measured at scale *s* for grid cell *i*

$Y_s$ – number of Western people measured at scale *s* for the whole urban area (core, hinterlands, FUA)

$_xP_{xs}$ – isolation index at scale *s*

$t_{si}$ – total population measured at scale *s* for grid cell *i*

$a_{si}$ – area measured at scale *s* for grid cell *i*

$A_s$ – total area measured at scale *s* for the whole urban area (core, hinterlands, FUA)

$P_{sxx}$, $P_{syy}$ – average proximity between members of the same group

$P_{sxy}$ – average proximity between members of different groups

$c_{ij}$ – negative exponential distance between cells *i* and *j*

$x_{sj}$ – number of non-Western people measured at scale *s* for grid cell *j*

$y_{sj}$ – number of Western people measured at scale *s* for grid cell *j*

$SP_s$ – spatial proximity index at scale *s*

$T_s$ – total population measured at scale *s* for the whole urban area (core, hinterlands, FUA)

$P_{stt}$ – average proximity among the total population

Table 1: Five dimensions of segregation, adapted from [Massey and Denton (1988)](), with an addition of the spatial scale

Hansen, R. (2003). Migration to Europe since 1945: Its history and its lessons. *The Political Quarterly, 74*, 25-38.

Hipp, J. R., & Boessen, A. (2013). Egohoods as waves washing across the city: a new measure of "neighborhoods". *Criminology, 51*(2), 287-327.

Johnston, R., Jones, K., Manley, D., & Owen, D. (2016). Macro-scale stability with micro-scale diversity: modelling changing ethnic minority residential segregation–London 2001–2011. Transactions of the Institute of British Geographers, 41(4), 389-402.

Krupka, D. J. (2007). Are big cities more segregated? Neighbourhood scale and the measurement of segregation. *Urban Studies, 44*(1), 187-197.

Leal, J., & Sorando, D. (2015). Economic crisis, social change and segregation processes in Madrid. Socio-Economic Segregation in European Capital Cities: East Meets West, 214-237.

Malheiros, J. M. 2002. Ethni-cities: residential patterns in Northern-European and Mediterranean metropolis. Implication in policy design. International Journal of Population Geography, 8: 107–134.

Manley, D., Johnston, R., Jones, K., & Owen, D. (2015). Macro-, meso-and microscale segregation: Modeling changing ethnic residential patterns in Auckland, New Zealand, 2001–2013. *Annals of the Association of American Geographers, 105*(5), 951-967.

Marcińczak, S., Mooses, V., Strömgren, M., & Tammaru, T. (2021). A comparative study of immigrant-native segregation at multiple spatial scales in urban Europe. Journal of Ethnic and Migration Studies, 1-23.

Massey, D. S., & Denton, N. A. (1988). The dimensions of residential segregation. *Social Forces, 67*(2), 281-315.

O'Sullivan, D., & Wong, D. W. (2007). A Surface-Based Approach to Measuring Spatial Segregation. *Geographical analysis, 39*(2), 147-168.

OECD. (2012). *Redefining Urban: A New Way to Measure Metropolitan Areas*: Organisation for Economic Cooperation and Development (OECD).

Östh, J., Clark, W. A. V., & Malmberg, B. (2014). Measuring the Scale of Segregation Using k-Nearest Neighbor Aggregates. *Geographical analysis, 47*(1), 34-49.
30